\documentclass[12pt]{article}
\usepackage[dvips]{graphicx}
\usepackage[american]{babel}
\usepackage{amsmath}
\newtheorem{theorem}{\bfseries Theorem}

\begin{document}
\title{Randomly evolving trees II}
\author{L. P\'al, \\\footnotesize{KFKI Atomic Energy Research Institute
1525 Budapest, P.O.B. 49. Hungary}}
\date{November 5, 2002}

\maketitle

\begin{abstract}
Generating function equation has been derived for the probability
distribution of the number of nodes with $k \geq 0$ outgoing lines
in randomly evolving special trees defined in an earlier paper
arXiv:cond-mat/0205650. The stochastic properties of the end-nodes
$(k=0)$ have been analyzed, and it was shown that the relative
variance of the number of end-nodes vs. time has a maximum when
the evolution is either subcritical or supercritical. On the
contrary, the time dependence of the relative dispersion of the
number of dead end-nodes shows a minimum at the beginning of the
evolution independently of its type. For the sake of better
understanding of the evolution dynamics the survival probability
of random trees has been investigated, and asymptotic expressions
have been derived for this probability in the cases of
subcritical, critical and supercritical evolutions. In critical
evolution it was shown that the probability to find the tree
lifetime larger than $x$, is decreasing to zero as $1/x$, if $x
\rightarrow \infty$. Approaching the critical state it has been
found the fluctuations of the tree lifetime to become extremely
large, and so near the critical state the average lifetime could
be hardly used for the characterization of the process.

\vspace{0.2cm}


\end{abstract}

\section{Introduction}

In a previous paper \cite{lpal02} we defined and analyzed random
processes with continuous time parameter describing the evolution
of special trees consisting of living and dead nodes connected by
lines. The initial state ${\mathcal S}_{0}$ of the tree
corresponds to a single living node called root which at the end
of its life capable to produce new living nodes, and after that it
becomes immediately dead. The new nodes are promptly connected to
the dead node and each of them  {\em independently of the others}
can evolve further like a root. It is evident that the random
evolution of this type is nothing else than a branching process.

In what follows we will use the notations applied in
\cite{lpal02}. Therefore, the distribution function of the
lifetime $\tau$ of a living node will be denoted by $T(t)$, and
the probability that the number $\nu$ of living nodes produced by
one dying precursor is equal to $j$ by $f_{j}$ where $j \in
{\mathcal Z}$.~\footnote{${\mathcal Z}$ is the set of non-negative
integers.}

In order to characterize the tree evolution two non-negative
integer valued random functions $\mu_{\ell}(t)$ and $\mu_{d}(t)$
were introduced in \cite{lpal02}. $\mu_{\ell}(t)$ is the number of
living nodes, while $\mu_{d}(t)$ is that of dead nodes at $t \geq
0$. It was mentioned also that the trivial equality
\[ \mu_{\ell}(t) + \mu_{d}(t) = \mu_{e}(t) + 1 \]
must be valid with probability $1$ at any $t \geq 0$. Here
$\mu_{e}(t)$ is the number of lines in the tree at the time moment
$t \geq 0$. For the generating functions
\begin{equation} \label{1}
g^{(\ell)}(t, z) = \sum_{n=0}^{\infty} p^{(\ell)}(t, n)\; z^{n},
\;\;\;\; \mbox{\'es} \;\;\;\; g^{(d)}(t, z) = \sum_{n=0}^{\infty}
p^{(d)}(t, n)\; z^{n},
\end{equation}
where
\[ p^{(\ell)}(t, n) = {\mathcal P}\{\mu_{\ell}(t)=n|{\mathcal S}_{0}\}
\;\;\;\; \mbox{\'es} \;\;\;\; p^{(d)}(t, n) = {\mathcal
P}\{\mu_{d}(t)=n|{\mathcal S}_{0}\} \] the following integral
equations were derived:
\begin{equation} \label{2}
g^{(\ell)}(t, z)= [1 - T(t)]\;z + \int_{0}^{t}
q\left[g^{(\ell)}(t-t', z)\right]\;dT(t')
\end{equation}
and
\begin{equation} \label{3}
g^{(d)}(t, z)= 1 -T(t) + z\;\int_{0}^{t} q\left[g^{(d)}(t-t',
z)\right]\;dT(t'),
\end{equation}
where
\[ q(z) = \sum_{j=0}^{\infty} f_{j}\;z^{j}.\]
If $T(t) = 1 - e^{-Qt}$, then the evolution becomes a Markov
process.

The nodes can be sorted into groups according to the number of
outgoing lines. Denote by $\mu(t, k)$ the number of nodes with $k
\geq 0$ outgoing lines at the time instant $t \geq 0$. A node not
having outgoing line is called {\em end-node}. It is obvious that
an end-node could be either live or dead. Therefore, the number of
end-nodes $\mu(t, 0)$ can be written as a sum of numbers of living
and dead end-nodes, i.e.
\[ \mu(t, 0) = \mu_{\ell}(t, 0) + \mu_{d}(t, 0). \]
Since all living nodes are end-nodes $\mu_{\ell}(t, 0)$ can be
replaced by $\mu_{\ell}(t)$. The total number of dead nodes
$\mu_{d}(t)$ is given by
\[ \mu_{d}(t) = \sum_{k=0}^{\infty} \mu_{d}(t, k). \]

In what follows we will calculate the probability distribution of
$\mu_{d}(t, k)$ and investigate the properties of end-nodes
$\mu(t, 0)$ playing important role in random tree evolution.

In order to have a deeper insight into the dynamics of the
evolution process, we will derive an important equation
determining the probability distribution function of the {\em tree
lifetime}.

\section{Generating functions}

\subsection{Distribution of $\nu$}

The basic properties of the probability distribution of the number
of nodes in a random tree are depending mainly on the distribution
law of the number $\nu$ of living nodes produced by one dying
precursor. In the sequel  we will use the notations
\[ {\bf E}\{\nu\} = q_{1} \;\;\;\;\;\; \mbox{and} \;\;\;\;\;\;
{\bf D}^{2}\{\nu\} = q_{2} + q_{1} - q_{1}^{2} \] introduced
already in \cite{lpal02} for the expectation and the variance of
$\nu$ where
\[ q_{j} =  \left[\frac{d^{j} q(z)}{dz^{j}}\right]_{z=1},
\;\;\;\;\;\; j = 1, 2, \ldots \] are the factorial moments of
$\nu$. It was shown in \cite{lpal02} that the time dependence of
the random evolution is determined almost completely by the
expectation value $q_{1}$. The evolution is called subcritical if
$q_{1}<1$, critical if $q_{1} = 1$ and supercritical if $q_{1}>1$.
In the further considerations we are going to use four simple
distributions for the random variable $\nu$.

\subsubsection{Arbitrary distribution}

It has been shown in \cite{lpal02} that the equations derived for
the first and the second moments of $\mu_{\ell}(t), \mu_{d}(t),
\ldots $ are true for any distribution of $\nu$ provided that the
moments $q_{1}$ and $q_{2}$ are finite. This type of distributions
of $\nu$ is called {\em arbitrary} and will be denoted by {\bf a}.

Many times it is expedient to assume distributions of $\nu$ to be
completely determined by one or two parameters.

\subsubsection{Geometric and Poisson distributions}

As known the geometric and Poisson distributions are containing
one parameter only, and so we have
\[ {\mathcal P}\{\nu = j\} = \left\{ \begin{array}{ll}
\frac{1}{1+q_{1}}\;\left(\frac{q_{1}}{1+q_{1}}\right)^{j}, &
\mbox{if $\nu \in {\bf g},$} \\
\mbox{ } & \mbox{ } \\
e^{-q_{1}}\;\frac{q_{1}^{j}}{j!}, & \mbox{if $\nu \in {\bf p},$}
\end{array} \right. \] where  {\bf g} and {\bf p} refer to the
geometric and the Poisson distributions, respectively. For the
sake of completeness we write
\[ q(z) = \left\{ \begin{array}{ll}
\frac{1}{1+(1-z)q_{1}}, & \mbox{if $\nu \in {\bf g}$,} \\
\mbox{ } & \mbox{ } \\
e^{-(1-z)q_{1}}, & \mbox{if $\nu \in {\bf p}$,}
\end{array} \right. \]
and
\[ {\bf E}\{\nu\} = \left\{ \begin{array}{ll}
q_{1}, & \mbox{if $\nu \in {\bf g}$,} \\
\mbox{ } & \mbox{ } \\
q_{1}, & \mbox{if  $\nu \in {\bf p}$,}
\end{array} \right.
\;\;\;\;\;\; \mbox{while} \;\;\;\;\;\; {\bf D}^{2}\{\nu\} =
\left\{ \begin{array}{ll}
q_{1}(1+q_{1}), & \mbox{if $\nu \in {\bf g}$,} \\
\mbox{ } & \mbox{ } \\
q_{1}, & \mbox{if $\nu \in {\bf p}$.} \end{array} \right.
 \]

\subsubsection{Truncated arbitrary distribution}

In this case the possible values of the random variable $\nu$ are
$0, 1$ and $2$ with probabilities $f_{0}, f_{1}$ and $f_{2}$,
respectively. This distribution will be denoted by {\bf t}. The
corresponding generating function of $\nu$ is given by
\[ q(z) = f_{0} + f_{1} z + f_{2} z^{2} = 1 + q_{1} (z-1) +
\frac{1}{2} q_{2} (z-1)^2. \]  This choice of $q(z)$ has a great
advantage, it makes possible to obtain exact solutions of the
generating function equations of $\mu_{\ell}(t), \mu_{d}(t),
\ldots$ characterizing the tree evolution.

\begin{figure} [ht!]
\protect \centering{\includegraphics[height=8cm,
width=10cm]{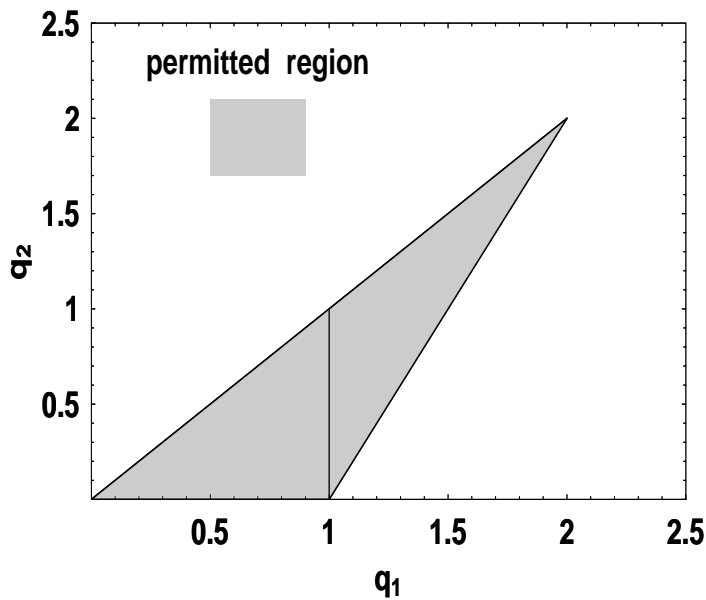}}\protect \vskip 0.2cm \protect
\caption{\footnotesize{The permitted values of $q_{1}$ and $q_{2}$
in the case of distribution {\bfseries t} of $\nu$.}} \label{fig1}
\end{figure}

It seems to be useful to cite the following trivial relations:
\[ f_{0} = 1 - q_{1} + \frac{1}{2}\;q_{2}, \;\;\;\;
f_{1} = q_{1} - q_{2},  \;\;\;\;  f_{2} = \frac{1}{2}\;q_{2}, \]
which follow from the equations
\[ f_{0} + f_{1} + f_{2}  =  1, \;\;\;\; f_{1} + 2f_{2} = q_{1},
\;\;\;\; 2f_{2} = q_{2}, \] where $q_{2} = {\bf D}^{2}\{\nu\} -
q_{1} + q_{1}^{2}$. Since $f_{0}, f_{1}$  and  $f_{2}$ are
non-negative, smaller than $1$, real numbers and their sum is
equal to $1$, the possible values of $q_{1}$ and $q_{2}$ are
restricted. The permitted values of $q_{1}$ and $q_{2}$ are shown
in Fig. \ref{fig1}. (See the shaded triangle!)

\subsection{Distribution of the number of dead
nodes with $k$ outgoing lines}

As has been already mentioned, a living node may create $k \geq 0$
new living nodes and after that it becomes immediately dead. This
node is called {\em dead node of out-degree $k$}. It was
introduced the random function $\mu_{d}(t,k)$ giving the number of
these dead nodes at time instant $t \geq 0$. Now, we want to
determine the probability that $\mu_{d}(t,k)$ is equal to $n \geq
0$ provided that at $t=0$ the tree was in the state ${\mathcal
S}_{0}$. First of all, we define the probability generating
function
\begin{equation} \label{4}
g_{k}^{(d)}(t, z) = \sum_{n=0}^{\infty} p_{k}^{(d)}(t, n)\;z^{n},
\end{equation}
where
\[ p_{k}^{(d)}(t, n) = {\mathcal P}\{\mu_{d}(t, k)=n|{\mathcal
S}_{0}\} \] is the probability that at time $t \geq 0$ the tree
has exactly $n$ dead nodes with $k \geq 0$ outgoing lines provided
that at $t=0$ it was in its initial state ${\mathcal S}_{0}$. By
using similar considerations as we did in \cite{lpal02}, we have
\[ p_{k}^{(d)}(t, n) = e^{-Qt}\;\delta_{n,0} + \]
\begin{equation} \label{5}
Q \int_{0}^{\infty} e^{-Q(t-t')} \left\{f_{0}\left[\delta_{0,k}
\delta_{n,1} + (1 - \delta_{0,k}) \delta_{n,0}\right] +
\sum_{j=1}^{\infty}f_{j} R_{j,k}^{(d)}(t', n)\right\}\;dt',
\end{equation}
where
\[ R_{j,k}^{(d)}(t', n) = \] \[\delta_{j,k}\;\sum_{n_{1}+\cdots+n_{j}=n-1}\;
\prod_{i=1}^{j}p_{k}^{(d)}(t', n_{i}) + (1 -
\delta_{j,k})\;\sum_{n_{1}+\cdots+n_{j}=n}\;\prod_{i=1}^{j}p_{k}^{(d)}(t',
n_{i}).\] The expression in square brackets at $f_{0}$ reflects
that two mutually excluding possibilities exist depending on
whether $k = 0$ or $k \neq 0$. One can see immediately that the
generating function defined by Eq. (\ref{4}) satisfies the
equation
\[ g_{k}^{(d)}(t, z) = e^{-Qt} + Q\;\int_{0}^{t} e^{-Q(t-t')}f_{0}[1 - (1-z)
\;\delta_{k,0}]\;dt' + \]  \[ Q \int_{0}^{t} e^{-Q(t-t')}
\left\{\sum_{j=1}^{\infty}f_{j}(1-\delta_{j,k})\left[g_{k}^{(d)}(t',z)
\right]^{j} + \sum_{j=1}^{\infty} f_{j}\delta_{j,k} z
\left[g_{k}^{(d)}(t', z)\right]^{j}\right\}\;dt'.\] By rearranging
the right side we find
\[ g_{k}^{(d)}(t, z) = e^{-Qt} - (1 - z)f_{0}\;\delta_{k,0}\;Q\;\int_{0}^{t}
 e^{-Q(t-t')}\;dt' - \]
\begin{equation} \label{6}
Q\;\int_{0}^{t} e^{-Q(t-t')}\left\{f_{k}\;(1 -z)\;(1
-\delta_{0,k})\;\left[ g_{k}^{(d)}(t',z) \right]^{k} -
q\left[g_{k}^{(d)}(t',z)\right]\right\}\;dt'.
\end{equation}
When $k > 0$ then the equation (\ref{6}) takes the form:
\[ g_{k}^{(d)}(t, z) = e^{-Qt} + \]
\begin{equation} \label{7}
Q\;\int_{0}^{t}
e^{-Q(t-t')}\left\{-f_{k}(1-z)\;\left[g_{k}^{(d)}(t',z)\right]^{k}
+ q\left[g_{k}^{(d)}(t', z)\right] \right\}\;dt'.
\end{equation}
The differential equation equivalent to (\ref{6}) is nothing else
than
\[ \frac{\partial g_{k}^{(d)}(t, z)}{\partial t} = - Q\;g_{k}^{(d)}(t, z) -
Q\;f_{0} (1 -z)\; \delta_{k,0} -  \]
\begin{equation} \label{8}
Q\;\left\{ f_{k}\;(1 -z)\;(1 -\delta_{0,k})\;
\left[g_{k}^{(d)}(t,z) \right]^{k} - q\left[g_{k}^{(d)}(t,
z)\right]\right\}
\end{equation}
and the initial condition is given by $ \lim_{t \downarrow 0}
g_{k}^{(d)}(t, z) = 1. $

\subsection{Joint distribution of the numbers of two dead \\ nodes
with different out-degrees}

The following step in our considerations is the determination of
the joint probability distribution
\begin{equation} \label{9}
{\mathcal P}\{\mu_d(t, k_{1})=n_{1},
\mu_d(t,k_{2})=n_{2}|{\mathcal S}_{0}\} = p_{k_{1},k_{2}}^{(d)}(t,
n_{1}, n_{2}),
\end{equation}
where $k_{1} \neq k_{2}$. It is clear from the definition
(\ref{9}) that $p_{k_{1},k_{2}}^{(d)}(t, n_{1}, n_{2})$ is the
probability that in the time interval $(0, t)$ the evolution
produces $\;n_{1}$ nodes with $k_{1}$ and $\;n_{2}$ nodes with
$k_{2}$ outgoing lines provided that at the moment $t=0$ the tree
was in its initial state ${\mathcal S}_{0}$. By using similar
arguments applied in deriving the backward equation (\ref{6}), we
obtain for the generating function
\begin{equation} \label{10}
g_{k_{1},k_{2}}^{(d)}(t, z_{1}, z_{2}) = \sum_{n_{1}=0}^{\infty}\;
\sum_{n_{2}=0}^{\infty} p_{k_{1},k_{2}}^{(d)}(t, n_{1},
n_{2})\;z_{1}^{n_{1}}\;z_{2}^{n_{2}}
\end{equation}
the following equation when $k_{1} \not= k_{2}$:
\[ g_{k_{1},k_{2}}^{(d)}(t, z_{1}, z_{2}) =
e^{-Qt} - Q\;f_{0}\;\int_{0}^{t} e^{-Q(t-t')}
[(1 - z_{1})\;\delta_{k_{1},0} + (1 -
z_{2})\;\delta_{k_{2},0}]\;dt' - \]
\[Q\;\int_{0}^{t} e^{-Q(t-t')}(1 - z_{1})\;f_{k_{1}}\;
\left[g_{k_{1},k_{2}}^{(d)}(t', z_{1}, z_{2})\right]^{k_{1}}\;dt'
- \]
\[ Q\;\int_{0}^{t} e^{-Q(t-t')}(1 -
z_{2})\;f_{k_{2}}\;\left[g_{k_{1},k_{2}}^{(d)}(t', z_{1},
z_{2})\right]^{k_{2}}\;dt' + \]
\begin{equation} \label{11}
Q\;\int_{0}^{t} e^{-Q(t-t')}q[g_{k_{1},k_{2}}^{(d)}(t', z_{1},
z_{2})]\;dt'.
\end{equation}
It follows from this equation that
\[ \lim_{z_{2} \rightarrow 0} g_{k_{1},k_{2}}^{(d)}(t, z_{1},
z_{2}) = g_{k_{1}}^{(d)}(t, z_{1}) \;\;\;\;\;\;\; \mbox{and}
\;\;\;\;\;\; \lim_{z_{1} \rightarrow 0} g_{k_{1},k_{2}}^{(d)}(t,
z_{1}, z_{2})= g_{k_{2}}^{(d)}(t, z_{2}). \] For the sake of the
completeness we are giving here the differential equation
equivalent to (\ref{11}). It has the form:
\[ \frac{\partial g_{k_{1},k_{2}}^{(d)}(t, z_{1}, z_{2})}{\partial t} =
-Q\;g_{k_{1},k_{2}}^{(d)}(t, z_{1}, z_{2}) +
Q\;q[g_{k_{1},k_{2}}^{(d)}(t, z_{1}, z_{2})] - \] \[ Q\;f_{0}[(1 -
z_{1})\;\delta_{k_{1},0} + (1 - z_{2})\;\delta_{k_{2},0}] - \]
\begin{equation} \label{12}
(1 - z_{1})\;f_{k_{1}}\;\left[g_{k_{1},k_{2}}^{(d)}(t, z_{1},
z_{2})\right]^{k_{1}} - (1 -
z_{2})\;f_{k_{2}}\;\left[g_{k_{1},k_{2}}^{(d)}(t, z_{1},
z_{2})\right]^{k_{2}},
\end{equation}
with the initial condition
\[ \lim_{t \downarrow 0} g_{k_{1},k_{2}}^{(d)}(t, z_{1}, z_{2}] = 1. \]

\subsection{Distribution of the number of end-nodes}

The dynamics of the random tree evolution is controlled by the the
end-nodes. It is evident that the living end-nodes are responsible
for the development of a tree, while the dead end-nodes represent
those points where the evolution was stopped. The probability
distribution of the number of dead end-nodes $p_{0}^{(d)}(t, n)$
can be obtained by substitution $k=0$ into Eq. (\ref{5}). It is
easy to show that the generating function
\[ g_{0}^{(d)}(t, z) = \sum_{n=0}^{\infty}p_{0}^{(d)}(t, n)\;z^{n} \]
satisfies the equation
\begin{equation} \label{13}
g_{0}^{(d)}(t, z) = e^{-Qt} - (1 - z) f_{0}(1 - e^{-Qt}) +
Q\;\int_{0}^{t} e^{-Q(t-t')}q\left[g_{0}^{(d)}(t', z)\right]\;dt'.
\end{equation}

In order to have an insight into the interplay between the living
and dead end-nodes it seems to be useful to calculate the
probability distribution of the random function
\[ \mu(t, 0) = \mu_{\ell}(t, 0) + \mu_{d}(t, 0) \]
and the joint distribution of $\mu_{\ell}(t, 0)$ and $\mu_{d}(t,
0)$.

With help of similar arguments used for the derivation of Eq.
(\ref{7}) it can be easily obtained the equation
\begin{equation} \label{14}
g_{0}(t, z) = e^{-Qt}\;z - (1 - z) f_{0}(1 - e^{-Qt}) +
Q\;\int_{0}^{t} e^{-Q(t-t')}q\left[g_{0}(t', z)\right]\;dt',
\end{equation}
for the generating function
\[ g_{0}(t, z) = \sum_{n=0}^{\infty} {\mathcal P}\{\mu(t,
0)=n|{\mathcal S}_{0}\}\;z^{n} = \sum_{n=0}^{\infty} p_{0}(t,
n)\;z^{n}, \] where $p_{0}(t, n)$ is the probability that the
number of all end-nodes at $t \geq 0$ is equal to $n$ provided
that at $t=0$ the tree was in the state ${\mathcal S}_{0}$.

Now we would like to  determine the joint distribution of
$\mu_{\ell}(t, 0)$ and $\mu_{d}(t,0)$, i.e. the probability
\[ {\mathcal P}\{\mu_{\ell}(t)=n_{1},\; \mu_{d}(t, 0)=n_{2}|{\mathcal S}_{0}\} =
p_{0}^{(\ell,d)}(t, n_{1}, n_{2}). \] It is evident that
\[ p_{0}^{(\ell,d)}(t, n_{1}, n_{2}) = e^{-Qt} \delta_{n_{1},1} \delta_{n_{2},0}
+ \] \[ + Q \int_{0}^{t} e^{-Q(t-t')}\;\left[f_{0}\delta_{n_{1},0}
\delta_{n_{2},1} + \sum_{k=1}^{\infty} f_{k}\;R_{0}^{(\ell,d)}(t',
n_{1}, n_{2})\right]\;dt', \] where
\[ R_{0}^{(\ell,d)}(t', n_{1}, n_{2}) =
\sum_{n_{11}+\cdots+n_{1k}=n_{1}}\;\sum_{n_{21}+\cdots+n_{2k}=n_{2}}\;
\prod_{j=1}^{k}p_{0}^{(\ell,d)}(t', n_{1j}, n_{2j}).  \] Simple
calculations show that the generating function
\[ g_{0}^{(\ell,d)}(t, z_{1}, z_{2}) = \sum_{n_{1}=0}^{\infty}\;
\sum_{n_{2}=0}^{\infty} p_{0}^{(\ell,d)}(t, n_{1},
n_{2})\;z_{1}^{n_{1}}\;z_{2}^{n_{2}} \] satisfies the equation
\[ g_{0}^{(\ell,d)}(t, z_{1}, z_{2}) = e^{-Qt}z_{1} - f_{0}(1 -
z_{2})(1 - e^{-Qt}) + \]
\begin{equation} \label{15}
+ Q \int_{0}^{t}e^{-Q(t-t')}\; q\left[ g_{0}^{(\ell,d)}(t, z_{1},
z_{2})\right]\;dt',
\end{equation}
which will be used in the next section for the determination of
the correlation  between the random variables $\mu_{\ell}(t, 0)$
and $\mu_{d}(t, 0)$.

Finally, we would like to cite the result of calculations
concerning the probability to find the random function
\[ \mu_{d}^{(o)}(t) = \sum_{k=1}^{\infty} \mu_{d}(t, k), \]
i.e. {\em the number of non end-nodes} at $t \geq 0$ to be equal
to $n$, provided that at $t=0$ the tree was in the state
${\mathcal S}_{0}$. Denote this probability by $p_{d}^{(o)}(t, n)$
and let us introduce the generating function
\[ g_{d}^{(o)}(t, z) = \sum_{n=0}^{\infty}{\mathcal P}\{\mu_{d}^{(o)}(t)
= n|{\mathcal S}_{0}\}\;z^{n} = \sum_{n=0}^{\infty}p_{d}^{(o)}(t,
n)\;z^{n}.  \] It can be proven that it satisfies the integral
equation
\begin{equation} \label{16}
g_{d}^{(o)}(t, z) = e^{-Qt} + f_{0}(1 - z) (1 - e^{-Qt}) + z Q
\int_{0}^{t} e^{-Q(t-t')}\;q\left[g_{d}^{(o)}(t', z)\right]\;dt',
\end{equation}
which is equivalent to the differential equation
\[ \frac{\partial g_{d}^{(o)}(t, z)}{\partial t} = Q f_{0} (1 -z) -
g_{d}^{(o)}(t, z) + Q z q\left[g_{d}^{(o)}(t, z)\right] \] with
initial condition $\lim_{t\rightarrow 0}g_{d}^{(o)}(t, z) = 1$.

\subsection{Average characteristics}

It is difficult to find solutions of the generating function
equations even in those cases when the distribution of $\nu$ is
known and simple.~\footnote{As has been already mentioned the
truncated ({\bf t}) arbitrary distribution of $\nu$ is one of the
rare exceptions when the generating function equation can be
exactly solved.} Therefore, in this section we would like to deal
with the average properties of tree evolution, and will derive
equations for the expectation values and variances of the number
of nodes of different kind. In order to have an insight into the
character of the stochastic interplay between the numbers of the
living and dead end-nodes, we will investigate the time variation
of the correlation between these nodes.

\subsubsection{Dead nodes with $k$ outgoing lines}

Let the first task be the calculation of the time dependence of
the expectation value and variance of dead nodes with $k \geq 0$
outgoing lines. By using the relation
\begin{equation} \label{17} {\bf E}\{\mu_{d}(t, k)\} =
\left[\frac{\partial g_{k}^{(d)}(t, z)}{\partial z}\right]_{z=1} =
m_{1}^{(d)}(t, k),
\end{equation}
from the probability generating function (\ref{6}) we  obtain that
\begin{equation} \label{18}
m_{1}^{(d)}(t, k) = f_{k}\;(1 - e^{-Qt}) + q_{1}\;Q\;\int_{0}^{t}
e^{-Q(t-t')}\;m_{1}^{(d)}(t', k)\;dt'.
\end{equation}
The solution of this equation can be written in the form:
\begin{equation} \label{19}
m_{1}^{(d)}(t, k) = \left\{
\begin{array}{ll}
\frac{f_{k}}{1-q_{1}}\;\left[1 - e^{-(1-q_{1})Qt}\right], &
\mbox{if $q_1 \neq 1$, } \\
\mbox{ } & \mbox{ } \\
f_k\;Qt, & \mbox{if $q_1=1$.}
\end{array} \right.
\end{equation}
Since
\[ \sum_{k=0}^{\infty} \mu_{d}(t, k) = \mu_{d}(t), \]
it is obvious that
\[ \sum_{k=0}^{\infty}m_{1}^{(d)}(t, k) =  m_{1}^{(d)}(t) =
\left\{ \begin{array}{ll} \frac{1}{1-q_{1}}\;\left[1 -
e^{-(1-q_{1})Qt}\right], & \mbox{if $q_{1} \neq 1$,} \\
\mbox{ } & \mbox{ } \\ Qt, &  \mbox{if $q_{1} = 1$,}
\end{array} \right. \] what has been already
derived in \cite{lpal02}. By using this relation we can conclude
that
\begin{equation} \label{20}
\frac{m_{1}^{(d)}(t,k)}{m_{1}^{(d)}(t)} = f_{k}, \;\;\;\;\;\;
\forall k \in {\mathcal Z}.
\end{equation}

In order to calculate the variance ${\bf D}^{2}\{\mu_{d}(t, k)\}$
we need the second factorial moment $m_{2}^{(d)}(t, k)$. It can be
shown that $m_{2}^{(d)}(t, k)$ satisfies the integral equation
\[ m_{2}^{(d)}(t, k) = q_{1}\;Q\;\int_{0}^{t}
e^{-Q(t-t')}\;m_{2}^{(d)}(t', k)\;dt' + \]
\[ Q\;\int_{0}^{t} e^{-Q(t-t')}\;\left\{2k\;f_{k}\;m_{1}^{(d)}(t', k) +
q_{2}\;\left[m_{1}^{(d)}(t', k)\right]^{2}\right\}\;dt' \] the
solution of which can be written in the form:
\[m_{2}^{(d)}(t, k) = \frac{f_{k}^{2}}{(1-q_{1})^{2}}\left(2k
+ \frac{q_{2}}{1-q_{1}}\right)\;(1 - e^{-\alpha t}) - \]
\begin{equation} \label{21}
2\;\frac{f_{k}^{2}}{1-q_{1}}\left(2k +
\frac{q_{2}}{1-q_{1}}\right)\;Qt\;e^{-\alpha t} +
\frac{f_{k}^{2}}{(1-q_{1})^{3}}\;q_{2}\;e^{-\alpha t}\;(1 -
e^{-\alpha t}),
\end{equation}
for all $q_{1} \neq 1$. Here the notation
\[ \alpha = (1 - q_{1})\;Q  \]
has been used. If $q_{1} = 1$, then the solution has the form:
\begin{equation} \label{22}
m_{2}^{(d)}(t, k) = f_{k}^{2}\;(Qt)^{2}\;\left(k +
\frac{1}{3}\;q_{2}\;Qt\right).
\end{equation}
Taking into account that
\[ {\bf D}^{2}\{\mu_{d}(t, k)\} = m_{2}^{(d)}(t, k) +
m_{1}^{(d)}(t, k)\left[1 - m_{1}^{(d)}(t, k)\right], \] where
\[ m_{1}^{(d)}(t, k)\left[1 - m_{1}^{(d)}(t, k)\right] =  \] \[
\frac{f_{k}}{1-q_{1}}\left(1 - \frac{f_{k}}{1-q_{1}}\right)\;(1 -
e^{-\alpha t}) + \frac{f_{k}^{2}}{(1-q_{1})^{2}}\;e^{-\alpha t}(1
- e^{-\alpha t}), \]  we obtain the following expression for the
variance:
\[ {\bf D}^{2}\{\mu_{d}(t, k)\} = \]
\[\frac{f_{k}}{1-q_{1}}\left[1 + \frac{f_{k}}{1-q_{1}}\left(2k - 1
+ \frac{q_{2}}{1-q_{1}}\right)\right]\;(1 - e^{-\alpha t}) - \]
\[2\;\frac{f_{k}^{2}}{1-q_{1}}\left(2k +
\frac{q_{2}}{1-q_{1}}\right)\;Qt\;e^{-\alpha t} + \]
\begin{equation} \label{23}
\frac{f_{k}^{2}}{(1-q_{1})^2}\left(1 +
\frac{q_{2}}{1-q_{1}}\right)\;e^{-\alpha t}(1 - e^{-\alpha t}),
\end{equation}
\[ q_{1} \neq 1 \;\;\;\;\;\; \mbox{and} \;\;\;\;\;\;
\forall k \in {\mathcal Z}, \] where $q_{2}$ can be replaced by
${\bf D}^{2}\{\nu\} - q_{1}\;(1-q_{1})$. When the evolution is
critical, i.e. when $q_{1} = 1$, then we have
\begin{equation} \label{24} {\bf
D}^{2}\{\mu_{d}(t, k)\} = f_{k}\;Qt\;\left[1 +(k-1)\;f_{k}\;Qt +
\frac{1}{3}\;f_{k}\;{\bf D}^{2}\{\nu\}\;(Qt)^{2}\right],
\end{equation}
\[ \forall k \in {\mathcal Z}. \]

If the time $t$ is converging to $\infty$ the Eqs. (\ref{23}) and
(\ref{24}) show that the variance remains finite in the case of
subcritical evolution only. Introducing the notation
\[ 1 + \frac{q_{2}}{1-q_{1}} = \frac{{\bf D}^{2}\{\nu\} +
(1-q_{1})^{2}}{1 - q_{1}}, \]  we obtain immediately the formula
\[ \lim_{t \rightarrow \infty}{\bf D}^{2}\{\mu_{d}(t, k)\} = \]
\begin{equation} \label{25}
\frac{f_{k}}{1-q_{1}}\;\left[1 + \frac{f_{k}}{1-q_{1}}\left(2(k-1)
+ \frac{q_{2}}{1 - q_{1}}\right)\right], \;\;\;\;\;\; \forall k
\in {\mathcal Z},
\end{equation}
\[ \mbox{if} \;\;\;\;\;\; q_{1} < 1. \]

\subsubsection{Properties of end-nodes}

Now we would like to deal with some properties of end-nodes. The
time dependence of the expectation of the number of end-nodes is
one of the simplest characteristics of randomly evolving trees. It
can be calculated from Eq. (\ref{14}). By using the relation
\[ {\bf E}\{\mu(t, 0)\} = m_{1}(t, 0) = \left[\frac{\partial
g_{0}(t, z)}{\partial z}\right]_{z=1}, \] after some elementary
mathematics we have
\begin{equation} \label{26}
m_{1}(t, 0) = \left\{ \begin{array}{ll} \frac{f_{0}}{1-q_{1}} +
\frac{1-f_{0}-q_{1}}{1-q_{1}}\;e^{-(1-q_{1}) Qt}, & \mbox{if
$q_{1} \neq 1$,} \\
\mbox{ } & \mbox{ } \\
1 + f_{0}\;Qt, & \mbox{if $q_{1} = 1$.} \end{array} \right.
\end{equation}
It is interesting to note that the expectation value of the number
of dead end-nodes $m_{1}^{(d)}(t, 0)$ can be easily calculated
from (\ref{26}). By substituting $m_{1}^{(\ell)}(t, 0) =
e^{-(1-q_{1})Qt}$ into the equation \[ m_{1}^{(d)}(t, 0) =
m_{1}(t, 0) - m_{1}^{(\ell)}(t, 0), \] one gets the formula
\begin{equation} \label{27}
m_{1}^{(d)}(t, 0) = \left\{ \begin{array}{ll}
\frac{f_{0}}{1-q_{1}}\;\left(1 - e^{-(1-q_{1})Qt}\right), &
\mbox{if $q_{1} \neq 1$, } \\
\mbox{ } & \mbox{ } \\
f_{0}\;Qt, & \mbox{if $q_{1} = 1$}
\end{array} \right.
\end{equation}
which is the same what follows from Eq. (\ref{19}) when $k=0$.

\begin{figure} [ht!]
\protect \centering{
\includegraphics[height=8cm, width=12cm]{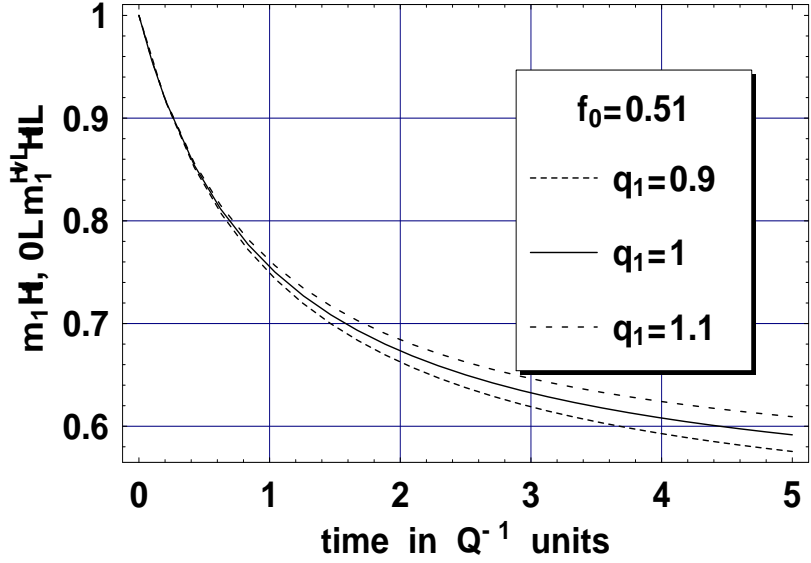}}\protect
\vskip 0.2cm \protect \caption{{\footnotesize Time dependence of
the ratio of the expected number of end-nodes $m_{1}(t, 0)$ to
that of all the nodes $m_{1}(t)$ in subcritical $(q_{1}=0.9)$,
critical $(q_{1}=1)$ and supercritical $(q_{1}=1.1)$ evolutions.}}
\label{fig2}
\end{figure}

In many cases it seems to be important to know how the evolution
process does alter the ratio of the expected number of end-nodes
$m_{1}(t, 0)$ to that of all the nodes $m_{1}(t)$. Taking into
account that~\footnote{See Eq. $(45)$ in \cite{lpal02}!}
\[ m_{1}(t) = \left\{ \begin{array}{ll}
\frac{1 - q_{1}\;e^{-(1-q_{1}) Qt}}{1 - q_{1}}, & \mbox{if $q_{1}
\neq 1$,} \\
\mbox{ } & \mbox{ } \\
1 + Qt, & \mbox{if $q_{1} = 1$,} \end{array} \right. \] after
elementary calculations we have
\[ \frac{m_{1}(t, 0)}{m_{1}(t)} = \left\{ \begin{array}{ll}
\frac{f_{0} + (1-f_{0}-q_{1})\;e^{-(1-q_{1}) Qt}}{1 -
q_{1}\;e^{-(1-q_{1}) Qt}}, & \mbox{if $q_{1} \neq 1$,} \\
\mbox{ } & \mbox{ } \\
\frac{1 + f_{0}\;Qt}{1 + Qt}, & \mbox{if $q_{1}=1$,} \end{array}
\right. \]  and Fig. \ref{fig2} shows the ratio vs. time plots for
values $q_{1} = 0.9, 1, 1.1$ at $f_{0}=0.51$. As seen the curves
are approaching the limit values
\[ \lim_{t \rightarrow \infty} \frac{m_{1}(t, 0)}{m_{1}(t)}
= \left\{ \begin{array}{ll} f_{0},  &  \mbox{if $q_{1} \leq 1$,}
\\ \mbox{ } & \mbox{ } \\
1 - \frac{1-f_{0}}{q_{1}}, & \mbox{if $q_{1} > 1$} \end{array}
\right. \] quite rapidly.\footnote{It is to note that $1 -
\frac{1-f_{0}}{q_{1}} > f_{0}$ if $q_{1} > 1$.}

In order to calculate the variance of the number of end-nodes we
need the second factorial moment of $\mu(t, 0)$. By using the
relation
\[ m_{2}(t, 0) = {\bf E}\{\mu(t, 0)\;[\mu(t, 0)-1]\} =
\left[\frac{\partial^{2} g_{0}(t, z)}{\partial z^{2}}\right]_{z=1}
\]
we obtain from Eq. (\ref{14}) the integral equation
\[ m_{2}(t, 0) = q_{1} Q\;\int_{0}^{t}e^{-Q(t-t')}\;m_{2}(t',
0)\;dt' + q_{2} Q\;\int_{0}^{t}e^{-Q(t-t')}\;\left[m_{1}(t',
0)\right]^{2}\;dt', \] the solution of that can be written in the
form
\[ m_{2}(t, 0) =
q_{2}\;\left(\frac{f_{0}}{1-q_{1}}\right)^{2}\;\frac{1 -
e^{-\alpha t}}{1 - q_{1}} + 2
q_{2}\;\frac{f_{0}(1-f_{0}-q_{1})}{(1-q_{1})^{2}} \;Qt\;e^{-\alpha
t} + \] \[ + q_{2}\;\left(\frac{1-f_{0}-q_{1}}{1-q_{1}}\right)^{2}
\;e^{-\alpha t}\;\frac{1 - e^{-\alpha t}}{1 - q_{1}}, \]
\[ \mbox{if}\;\;\;\;\; q_{1} \neq 1. \]
When $q_{1} = 1$, i.e. when the evolution is critical we get
\[ m_{2}(t, 0) = q_{2}\left[Qt + f_{0}\;(Qt)^{2} + \frac{1}{3}
f_{0}^{2}\;(Qt)^{3}\right]. \] Finally we have the variance of
$\mu(t, 0)$ in the following form:
\[ {\bf D}^{2}\{\mu(t, 0)\} =
\left[\frac{f_{0}(1-f_{0}-q_{1})}{(1-q_{1})^{2}} + \frac{q_{2}}{1
-q_{1}}\;\left(\frac{f_{0}}{1-q_{1}}\right)^{2}\right](1 -
e^{-\alpha t}) + \] \[ 2 q_{2}\frac{f_{0}(1-f_{0}-q_{1})}
{(1-q_{1})^{2}}\;Qt\;e^{-\alpha t} +  \]
\begin{equation} \label{28}
\left(1 + \frac{q_{2}}{1-q_{1}}\right)\;\left(\frac{1-f_{0}-q_{1}}
{1-q_{1}}\right)^{2}\;e^{-\alpha t}(1 - e^{-\alpha t}),
\end{equation}
\[ \mbox{if} \;\;\;\;\; q_{1} \neq 1,  \]
and
\begin{equation} \label{29}
{\bf D}^{2}\{\mu(t, 0)\} = (q_{2} - f_{0})\;(1 + f_{0}\;Qt)\;Qt +
\frac{1}{3} q_{2} f_{0}^{2}\;(Qt)^{3},
\end{equation}
\[ \mbox{if} \;\;\;\;\; q_{1} = 1.  \]

\vspace{0.4cm}

{\footnotesize It can be easily proven that if $q_{1}=1$, then
$q_{2} - f_{0}
> 0$. Since $q_{1} =
\sum_{k=1}^{\infty}k f_{k}=1$ and $\sum_{k=0}^{\infty} f_{k} = 1$
it is evident that
\[ f_{0} + f_{1} + \sum_{k=2}^{\infty} f_{k} = f_{1} +
\sum_{k=2}^{\infty} k f_{k}, \] and hence
\[ f_{0} = \sum_{k=2}^{\infty} (k-1)\;f_{k}. \] By using this
expression of $f_{0}$ we find that
\[ q_{2} - f_{0} = \sum_{k=2}^{\infty}[k(k-1) - (k-1)]\;f_{k} =
\sum_{k=2}^{\infty} (k-1)^{2}\;f_{k} >0. \;\;\;\;\;\;
\mbox{Q.E.D.} \]}

\vspace{0.4cm}

As follows from Eqs. (\ref{28}) and (\ref{29}) when $t \rightarrow
\infty$ then the variance of $\mu(t, 0)$ tends to finite limit
value in the case of subcritical evolution only,

\begin{figure} [ht!]
\protect \centering{
\includegraphics[height=12cm, width=10cm]{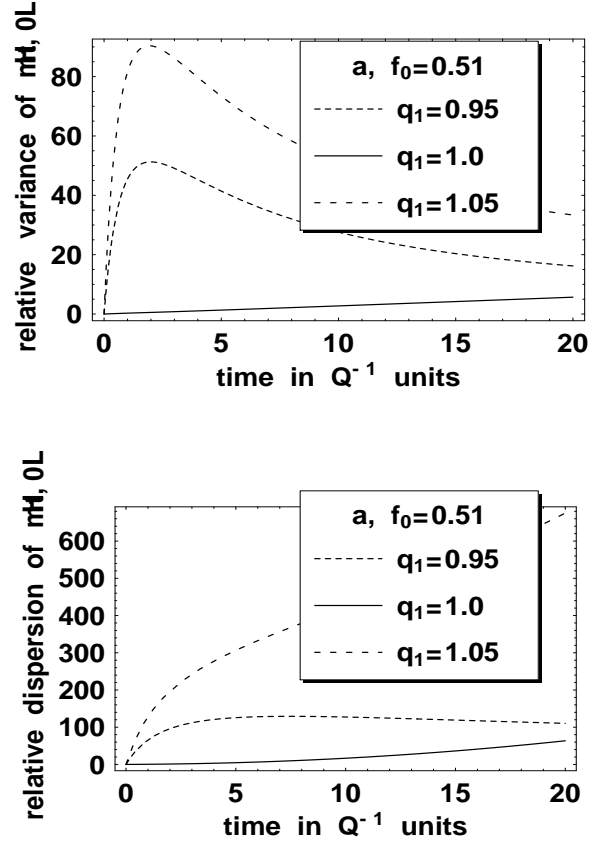}}\protect
\vskip 0.2cm \protect \caption{{\footnotesize Relative variance
(upper) and dispersion (lower) of the number of end-nodes vs. time
in subcritical, critical and supercritical trees. The distribution
of $\nu$ is arbitrary and ${\bf D}^{2}\{\nu\}=0.9$.}} \label{fig3}
\end{figure}

In order to show the main features of the time variation of
fluctuations occurring in the end-node number of trees the
relative variance and the relative dispersion of $\mu(t, 0)$ vs.
time have been calculated. The results are plotted in Fig.
\ref{fig3}. It has been assumed that the distribution of $\nu$ is
arbitrary in that sense as it was defined in section 2.1.1. In the
upper part of the Fig. \ref{fig3} one can see that the relative
variance of $\mu(t, 0)$ reaches a maximum just after the beginning
of the process but in the case of subcritical and supercritical
evolutions only. If $t \rightarrow \infty$, then we have
\[ \lim_{t \rightarrow \infty} \frac{{\bf D}^{2}\{\mu(t,
0)\}}{{\bf E}^{2}\{\mu(t, 0)\}} = \left\{ \begin{array}{ll}
\frac{1-f_{0}-q_{1}}{f_{0}} + \frac{q_{2}}{1-q_{1}}, & \mbox{if
$q_{1} < 1$,} \\
\mbox{ } & \mbox{ } \\
\frac{q_{2}}{q_{1}-1} - 1, & \mbox{if $q_{1} > 1$,} \end{array}
\right. \]

{\footnotesize The proof of the inequality $q_{2} \geq q_{1} - 1$
is very simple. If we substitute $q_{2}$ and $q_{1}$ by
$\sum_{k=1}^{\infty} k(k-1)\;f_{k}$ and  $\sum_{k=1}^{\infty}
k\;f_{k}$, respectively, then we can write
\[ \sum_{k=1}^{\infty}k(k-1)\;f_{k} - \sum_{k=1}^{\infty} k\;f_{k} + 1
\geq 0, \] i.e.
\[ \sum_{k=1}^{\infty}(k-1)^{2}\;f_{k} + f_{0} \geq 0, \]
and this is trivially true.}

\begin{figure} [ht!]
\protect \centering{
\includegraphics[height=12cm, width=10cm]{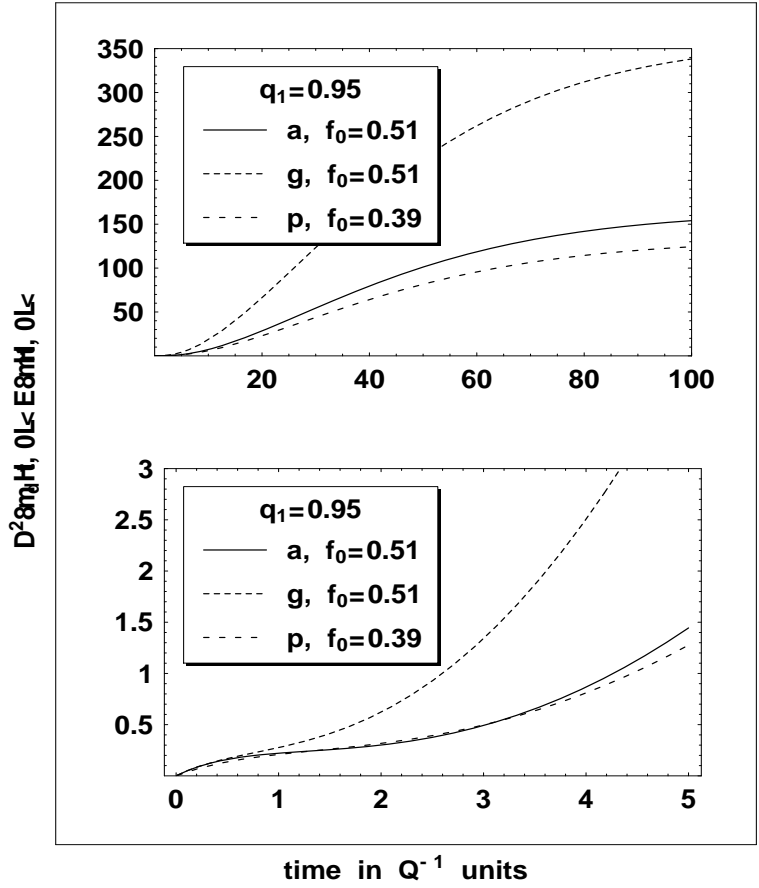}}\protect
\vskip 0.2cm \protect \caption{{\footnotesize Relative dispersion
of the number of end-nodes in subcritical trees vs. time. Curves
{\bf a, g}, and {\bf p} correspond to the arbitrary, geometric and
Poisson distributions of $\nu$, respectively. In the case of
arbitrary distribution: ${\bf D}^{2}\{\nu\}=0.9$. Upper: $0 \leq
Qt \leq 100$. Lower: $0 \leq Qt \leq 5$.}} \label{fig4}
\end{figure}
If the evolution is critical $(q_{1}=1)$, then the relative
variance
\[ \frac{{\bf D}^{2}\{\mu(t, 0)\}}{{\bf E}^{2}\{\mu(t, 0)\}} =
(q_{2} - f_{0})\;\frac{Qt}{1 + f_{0} Qt} + \frac{1}{3} q_{2}
f_{0}^{2}\;\frac{(Qt)^{3}}{(1 + f_{0} Qt)^{2}} \] is increasing
monotonously to infinity with $t$. The curves in the lower part of
Fig. \ref{fig3} show the time variation of the relative dispersion
of $\mu(t, 0)$ which has a finite limit value
\[ \lim_{t \rightarrow \infty} \frac{{\bf D}^{2}\{\mu(t,
0)\}}{{\bf E}\{\mu(t, 0)\}} = 1 +
\frac{f_{0}}{1-q_{1}}\left(\frac{q_{2}}{1-q_{1}} - 1\right) \] in
the the subcritical evolution only.

The influence of the distribution law of $\nu$ on the relative
dispersion can be seen in Fig. \ref{fig4}. The calculations have
been carried out in the case of subcritical evolution. In the
upper part of the figure it can be seen that each curves tends to
a finite asymptotic value when $t \Rightarrow \infty$. It is
remarkable that the geometric distribution of $\nu$ brings about
much larger fluctuations in $\mu(t, 0)$ than the arbitrary and
Poisson distributions. The lower part of the figure shows the
beginning of the time dependence which is reflecting the effect of
two competing processes. One of them is the formation of new
living nodes, while the other one is the death of end-nodes.

It seems to be worthwhile to calculate the time dependence of the
expectation and variance of the number of dead end-nodes
$\mu_{d}(t, 0)$. Substituting $k=0$ into the Eqs. (\ref{19}) and
(\ref{23}) it can be derived the relative variance
\[ \frac{{\bf D}^{2}\{\mu_{d}(t, 0)\}}{{\bf E}^{2}\{\mu_{d}(t,
0)\}} = \left(\frac{q_{2}}{1-q_{1}} + \frac{1-q_{1}}{f_{0}} -
1\right)\;\frac{1}{1 - e^{-\alpha t}} - \] \[ -
2q_{2}\;Qt\;\frac{e^{-\alpha t}}{(1 - e^{-\alpha t})^{2}} +
\left(1 + \frac{q_{2}}{1-q_{1}}\right)\;\frac{e^{-\alpha t}}{1 -
e^{-\alpha t}}, \]
\[ \mbox{if} \;\;\;\;\;\; q_{1} \neq 1,  \]
and
\[ \frac{{\bf D}^{2}\{\mu_{d}(t, 0)\}}{{\bf E}^{2}\{\mu_{d}(t,
0)\}} = \frac{1}{3} q_{2}\;Qt - 1 + \frac{1}{f_{0}\;Qt}, \]
\[  \mbox{if} \;\;\;\;\;\; q_{1} = 1.  \]

The relative variance of $\mu_{d}(t, 0)$ converges to finite
value, when $t \rightarrow \infty$, in both subcritical and
supercritical evolutions but diverges if the evolution is
critical. Let us introduce the notations
\[ \lim_{t \rightarrow \infty}\;\frac{{\bf D}^{2}\{\mu_{d}(t, 0)\}}
{{\bf E}^{2}\{\mu_{d}(t, 0)\}} = \left\{ \begin{array}{ll}
rv_{d}^{(a)}, & \mbox{if $\nu \in {\bf a}$,} \\
\mbox{ } & \mbox{ } \\
rv_{d}^{(g)}, & \mbox{if $\nu \in {\bf g}$,} \\
\mbox{ } & \mbox{ } \\
rv_{d}^{(p)}, & \mbox{if $\nu \in {\bf p}$,}
\end{array} \right. \]
and summarize the limit values of relative variances. We find that
\[ rv_{d}^{(a)} = \left\{
\begin{array}{ll}
\frac{1-q_{1}}{f_{0}} - 1- q_{1} + \frac{{\bf
D}^{2}\{\nu\}}{1-q_{1}}, & \mbox{if
$q_{1}<1$,} \\
\mbox{ } & \mbox{ } \\
\infty, & \mbox{if $q_{1}=1$,} \\
\mbox{ } & \mbox{ } \\
q_{1} - 1 + \frac{{\bf D}^{2}\{\nu\}}{q_{1}-1}, & \mbox{if
$q_{1}>1$.}
\end{array} \right. \]
If the distribution of $\nu$ is geometric, then
\[ rv_{d}^{(g)} =
\left\{ \begin{array}{ll}
q_{1}^{2}\;\frac{1+q_{1}}{1-q_{1}}, & \mbox{if $q_{1}<1$,} \\
\mbox{ } & \mbox{ } \\
\infty, & \mbox{if $q_{1}=1$,} \\
\mbox{ } & \mbox{ } \\
\frac{2q_{1}^{2}}{q_{1}-1} - 1, & \mbox{if $q_{1}>1$,}
\end{array} \right. \]
while if it is of Poisson type, then
\[ rv_{d}^{(p)} =
\left\{ \begin{array}{ll} (1-q_{1})\;e^{q_{1}} +
\frac{q_{1}^{2} + q_{1} -1}{1-q_{1}}, & \mbox{if $q_{1}<1$,} \\
\mbox{ } & \mbox{ } \\
\infty, & \mbox{if $q_{1}=1$,} \\
\mbox{ } & \mbox{ } \\
\frac{q_{1}^{2}}{q_{1}-1} - 1, & \mbox{if $q_{1}>1$.}
\end{array} \right. \]
\begin{figure} [ht!]
\protect \centering{
\includegraphics[height=11.3cm, width=8cm]{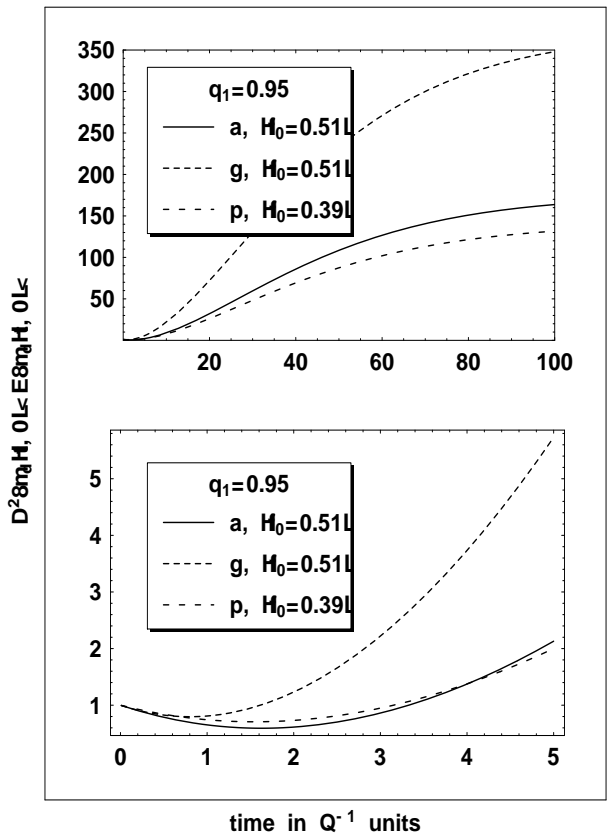}}\protect
\vskip 0.2cm \protect \caption{{\footnotesize Relative dispersion
of the number of dead end-nodes vs. time in subcritical trees.
Curves {\bf a, g}, and {\bf p} correspond to the arbitrary,
geometric and Poisson distributions of $\nu$, respectively. In the
case of arbitrary distribution: ${\bf D}^{2}\{\nu\}=0.9$ and
$f_{0}=0.51$. Upper: $0 \leq Qt \leq 100$. Lower: $0 \leq Qt \leq
5$.}} \label{fig5}
\end{figure}
Finally, it seems to be interesting to look at the time dependence
of the relative dispersion the number of dead end-nodes. It
follows from Eqs. (\ref{19}) and (\ref{23}) that
\[ \frac{{\bf D}^{2}\{\mu_{d}(t, 0)\}}{{\bf E}\{\mu_{d}(t, 0)\}} =
1 + \frac{f_{0}}{1-q_{1}}\;\left(\frac{q_{2}}{1-q_{1}} - 1\right)
+ \] \[ + \frac{f_{0}}{1-q_{1}}\;\left(\frac{q_{2}}{1-q_{1}} +
1\right) \; e^{-\alpha t} -
2f_{0}\;\frac{q_{2}}{1-q_{1}}\;Qt\;\frac{e^{-\alpha
t}}{1-e^{-\alpha t}}, \]
\[ \mbox{if} \;\;\;\;\;\; q_{1} \neq 1, \]
and
\[ \frac{{\bf D}^{2}\{\mu_{d}(t, 0)\}}{{\bf E}\{\mu_{d}(t, 0)\}} =
1 + f_{0}\;Qt\left(\frac{1}{3}q_{2}\;Qt - 1\right), \]
\[ \mbox{if} \;\;\;\;\;\; q_{1} = 1. \]
Fig. \ref{fig5} shows the time dependence of the relative
dispersion of the dead end-nodes $\mu_{d}(t, 0)$ in subcritical
evolution for all the three ({\bf a, g, p}) distributions of
$\nu$. In the lower part of the figure one can see that the
relative dispersion curves have minimum just after the beginning
of the process.(In the case of the arbitrary distribution of $\nu$
the minimum sites are given in Table I.)

\begin{center}
Table I: {\footnotesize Sites of the minimum in relative
dispersion vs. time curves at several $q_{1}$ and ${\bf
D}^{2}\{\nu\}$ values. (Time is measured in $Q^{-1}$ units and
$\nu \in {\bf a}$.)}

\vspace{0.2cm}

\begin{tabular}{|c|c|c|c|c|c|} \hline
{\footnotesize ${\bf D}^{2}\{\nu\} \backslash q_{1}$} &
 0.90 & 0.95 & 1.00 & 1.05 & 1.10 \\
\hline
  0.90   &  1.7724 & 1.7223 & 1.6617 & 1.6069 & 1.5441  \\ \hline
  1.00   &  1.5849 & 1.5448 & 1.5000 & 1.4514 & 1.3999  \\ \hline
  1.10   &  1.4332 & 1.4006 & 1.3636 & 1.3233 & 1.2803  \\ \hline
\end{tabular}
\end{center}

\section{Lifetime of trees}

\subsection{General considerations}

It is obvious that the evolution of a random tree will stop at
that time instant $\theta$ which satisfies with probability one
the equation $\mu_{\ell}(\theta)=0$. The random variable $\theta$
is called {\em lifetime of the tree}. In order to determine its
distribution function
\begin{equation} \label{30}
{\mathcal P}\{\theta \leq t|{\mathcal S}_{0}\}= L(t),
\end{equation}
one has to recognize that the probability \[ {\mathcal
P}\{\mu_{\ell}(t)=0|{\mathcal S}_{0}\} = p^{(\ell)}(t, 0)\] to
find zero living node at time moment $t \geq 0$ in a tree is the
same as the probability that the lifetime $\theta$ of that tree is
not larger than $t \geq 0$, therefore, one can write

\[ {\mathcal P}\{\theta \leq t|{\mathcal S}_{0}\} = {\mathcal
P}\{\mu_{\ell}(t)=0| {\mathcal S}_{0}\},\] i.e.,
\begin{equation} \label{31}
L(t) = p^{(\ell)}(t,0) = \lim_{z \downarrow 0}\; g^{(\ell)}(t, z).
\end{equation}
It is clear that if $0 < t_{1} \leq t_{2}$ then $L(t_{1}) \leq
L(t_{2})$, i.e., $L(t)$ is a non-decreasing function of its
argument, hence the limit relation
\begin{equation} \label{32}
\max_{0<t\leq \infty}\;L(t) = \lim_{t \rightarrow \infty}\; L(t) =
L_{\infty} \leq 1
\end{equation}
must be valid. We will call the quantity $L_{\infty}$ {\em
dying-out-probability}, and prove the following theorem:

\begin{theorem} \label{t1}
If $q_{1} \leq 1$, i.e., the random evolution is not
supercritical, then $L_{\infty}=1$, while if $q_{1} > 1$, i.e.,
the evolution is supercritical, then $L_{\infty}$ is equal to the
single, smaller than $1$, non-negative root of the function
$\psi(y) = q(y) - y, \;\; y \in [0, 1]$.~\footnote{It is evident
that $\psi(1) = 0$.}
\end{theorem}

{\bfseries Proof.} For the proof we exploit the fundamental
property of the generating function $g^{(\ell)}(t, z)$ which is
expressed by the equation
\[ g^{(\ell)}(t+u, z) = g^{(\ell)}\left[t, g^{(\ell)}(u,
z)\right]. \] Applying the relation (\ref{31}) we have
\[ L(t+u) = g^{(\ell)}\left[t, L(u)\right] \]
and since
\[ \lim_{u \rightarrow \infty}\;L(u) = \lim_{u \rightarrow
\infty}\;L(t+u)= L_{\infty}, \] we can write for every $t \geq 0$
that
\[ L_{\infty} = g^{(\ell)}(t, L_{\infty}). \] Putting
$g^{(\ell)}(t, L_{\infty})$ into
\begin{equation} \label{33}
\frac{dg^{(\ell)}(t,z)}{dt} = Q\;q\left[g^{(\ell)}(t,z)\right] -
Q\; g^{(\ell)}(t,z)
\end{equation}
derived from (\ref{2}), we obtain
\begin{equation} \label{34}
q\left(L_{\infty}\right) - L_{\infty} = 0.
\end{equation}
Considering that $q(y)$ is probability generating function, i.e.,
$\lim_{y \uparrow 1}\;q(y) = 1$, according to a well-known theorem
of generating functions \cite{lpal94}, it is clear that if $q_{1}
> 1$ then Eq. (\ref{34}) ---  besides the trivial fixed-point
$1$ ---  must have an other, smaller than $1$, non-negative
fixed-point $L_{\infty}$ too, and this is what we wanted to prove.
Q.E.D.~\footnote{The proof of this and the following theorems is
based on Sjewastjanow's ideas \cite{sewast74}.}

Let us introduce the probability
\begin{equation} \label{35}
S(x) = 1 - L(x),
\end{equation}
to find the tree at the time instant $x=Qt$ in living state. The
$S(x)$ will be called {\em survival probability}. By taking into
account the properties of $L(x)$ one obtains
\begin{equation} \label{36}
\lim_{Qt \rightarrow \infty} S(x) = \left\{ \begin{array}{ll}
         0, &  \mbox{if $q_{1} \leq 1$,} \\
         \mbox{} & \mbox{} \\
         S_{\infty}=1-L_{\infty}, & \mbox{if $q_{1} > 1$.}
         \end{array} \right.
\end{equation}
By using Eqs. (\ref{33}), (\ref{31}), and (\ref{35}) one gets
\begin{equation} \label{37}
\frac{dS}{dx} = -q(1-S) - S + 1,
\end{equation}
which has the solution
\begin{equation} \label{38}
x = \int_{S(x)}^{1} \frac{dy}{q(1-y) + y - 1},
\end{equation}
if the initial condition is $S(0)=1$. Now, we would like to derive
{\em asymptotic expressions} of $S(x)$ for large $x=Qt$ in the
cases of subcritical, critical and supercritical evolution.

\begin{theorem} \label{t2}
If the integral
\begin{equation} \label{39}
\int_{0}^{1}\frac{q(1-y) + q_{1}y - 1}{y[q(1-y) + y - 1]}\;dy =
-\log K
\end{equation}
is finite, then in the case of subcritical evolution the survival
probability $S(x)$ has the following form:
\begin{equation} \label{40}
S(x) = K\;e^{-(1-q_{1})x}\;[1 + o(1)]
\end{equation}
when $x=Qt \Rightarrow \infty$.
\end{theorem}

{\bfseries Proof.} The proof is simple: the identity
\[ \log \frac{e^{-(1-q_{1})x}}{S(x)} = (q_{1} - 1)\;x - \log S(x),
\] by using the Eq. (\ref{38}), can be rewritten in the form
\[ \log \frac{e^{-(1-q_{1})x}}{S(x)} = (q_{1}-1)\;\int_{S(x)}^{1}
\frac{dy}{q(1-y) + y - 1} + \int_{S(x)}^{1}\frac{dy}{y} = \]
\[ \int_{S(x)}^{1} \frac{q(y-1) + q_{1}\;y - 1}{y\;[q(1-y) + y -
1]}\;dy = k(x), \] and hence \[ S(x) = e^{-k(x)}\;e^{-(1-q_{1})x}.
\] From this we find
\[ \lim_{x \rightarrow \infty} \frac{S(x)}{e^{-(1-q_{1})x}} =
e^{-k(\infty)}, \] where
\[ k(\infty) = \int_{0}^{1}\frac{q(1-y) + q_{1}y - 1}
{y[q(1-y) + y - 1]}\;dy = - \log K \] since $S(\infty)=0$ if
$q_{1}<1$. According to the assumption (\ref{39}) we can write
\[ S(x) = e^{-k(\infty)}\;e^{-(1-q_{1})x}\;[1 + o(1)] =
K\;e^{-(1-q_{1})x}\;[1 + o(1)], \] and this is what we wanted to
prove. Q.E.D.

It is to mention that in the case of {\bfseries t} distribution of
$\nu$ from Eq. (\ref{39}) we obtain
\[ K = \frac{1}{1 + \frac{1}{2}\;\frac{q_{2}}{1 - q_{1}}}, \]
and so
\[ S(x) = \frac{e^{-(1-q_{1})x}}
{1 + \frac{1}{2}\;\frac{q_{2}}{1 - q_{1}}}\;[1 + o(1)]. \]

\begin{theorem} \label{t3}
If $q_{2} < \infty$ and $q_{1}=1$, then the asymptotic expression
for the survival probability is given by
\begin{equation} \label{41}
S(x) = \frac{2}{q_{2}\;x}\;[1 + o(1)]
\end{equation}
when $x=Qt \Rightarrow \infty$.
\end{theorem}

{\bfseries Proof.} By using the series expansion theorem according
to which
\[ q(1-S) = 1 - S + \frac{1}{2}\;q''[b(x)]\;S^{2}, \]
where $1-S \leq  b(x) < 1$, from Eq. (\ref{37}) we obtain
\[ \frac{dS}{dx} = - \frac{1}{2}\;q''[b(x)]\;S^{2}. \]
Since if $x \Rightarrow \infty$, then $S(x) \Rightarrow 0$ we can
write
\[ q''[b(x)] = q_{2} + \epsilon(x), \;\;\;\;\;\; \mbox{where}
\;\;\;\;\;\; \lim_{x \rightarrow \infty}\;\epsilon (x) = 0.\]
Taking into account this expression for $q''[b(x)]$ we have
\[ \frac{dS}{dx} = - \frac{1}{2}\;q_{2}\;S^{2} -
\frac{1}{2}\;\epsilon(x)\;S^{2}, \] the solution of which  can be
written in the form
\[ S(x) = \left[\frac{1}{2}\;q_{2}\;x +
\frac{1}{2}\;\int_{0}^{x}\epsilon(v)\;dv + C\right]^{-1}. \] The
initial condition $S(0) = 1$ results in $C=1$, and therefore
\[ S(x) = \frac{2}{q_{2}\;x}\;\left[1 + \frac{2}{q_{2} x} +
\frac{1}{q_{2} x}\;\int_{0}^{x}\epsilon(v)\;dv\right]^{-1}. \] By
applying the L'Hospital rule we find
\[ \lim_{x \rightarrow \infty}
\frac{1}{x}\;\int_{0}^{x}\epsilon(v)\;dv = \lim_{x \rightarrow
\infty} \epsilon(x) = 0, \] and hence
\[ S(x) = \frac{2}{q_{2}\;x}\;[1 + o(1)], \]
what we wanted to prove. Q.E.D.

The third task is to derive the asymptotic expression for $S(x)$
in the case of supercritical evolution.
\begin{theorem} \label{t4}
If the integral
\begin{equation} \label{42}
\int\limits_{S_{\infty}}^{1} \frac{q(1-y) + y -1 -(q_{1}-1)\;(y -
S_{\infty})}{(y - S_{\infty})\;[q(1-y) + y -1]}\;dy = r(\infty)
\end{equation}
is finite in the case of $q_{1} > 1$, then
\begin{equation} \label{43}
S(x) = S_{\infty} + (1 -
S_{\infty})\;e^{-r(\infty)}\;e^{-(q_{1}-1)x}\;[1 + o(1)],
\end{equation}
for $x = Qt \Rightarrow \infty$ where $S_{\infty} = 1 - L_{\infty}
< 1$ is the limit value of the survival probability.
\end{theorem}

{\bfseries Proof.} From the identity
\[ \log \frac{e^{-(q_{1}-1)x}}{S(x)-S_{\infty}} =
-(q_{1} - 1)\;x - \log [S(x) -S_{\infty}]\] we obtain
\begin{equation} \label{44}
\log \frac{e^{-(q_{1}-1)x}}{S(x)-S_{\infty}} = - (q_{1} -
1)\;\int\limits_{S(x)}^{1}\frac{dy}{q(1-y) + y - 1} +
\int\limits_{S(x)-S_{\infty}}^{1}\frac{dy}{y},
\end{equation}
and since
\[ \int\limits_{S(x)-S_{\infty}}^{1}\;\frac{dy}{y} =
\int\limits_{S(x)}^{1} \frac{dy}{y - S_{\infty}} +
\log\;\frac{1}{S(x)-S_{\infty}},  \] we can rewrite Eq. (\ref{44})
in the form
\[ \log \;\frac{1 - S_{\infty}}{S(x) -
S_{\infty}}\;e^{-(q_{1}-1)x} = r(x), \] where
\[ r(x) = \int\limits_{S(x)}^{1} \frac{q(1-y) + y -1 -(q_{1}-1)\;(y -
S_{\infty})}{(y - S_{\infty})\;[q(1-y) + y -1]}\;dy,  \] and so we
have
\[ S(x) = S_{\infty} + (1 -
S_{\infty})\;e^{-r(x)}\;e^{-(q_{1}-1)x}. \] It has been assumed
$r(\infty)$ to be finite and so
\[ \lim_{x \rightarrow \infty} \frac{S(x) -
S_{\infty}}{e^{-(q_{1}-1)x}} = (1 - S_{\infty})\;e^{-r(\infty)} <
\infty. \] Thus the theorem is proved. Q.E.D.

In the case of {\bfseries t} distribution of $\nu$ we find
immediately that
\[ S_{\infty} = 2 \frac{q_{1} - 1}{q_{2}}, \;\;\;\;\;\; \mbox{and} \;\;\;\;\;\;
r({\infty}) = - \log S_{\infty},  \] and hence obtain
\[ S(x) = 2 \frac{q_{1} - 1}{q_{2}} + 2 \frac{q_{1} - 1}{q_{2}}
\;\left(1 - 2 \frac{q_{1} - 1}{q_{2}}\right)\;e^{-(q_{1}-1)x}\;[1
+ o(1)],
\] when $x=Qt \Rightarrow \infty$.

\subsection{Some exactly solvable models}

In order to demonstrate the characteristic features of the
lifetime of random trees we will use such probabilities $f_{k},
\;\; k=0, 1, \ldots $ for the offspring numbers $\nu$ that make
possible to solve exactly the equation (\ref{37}) of the survival
probability. In the following, two special cases will be
investigated, namely
\begin{equation} \label{45}
q(z) = \left\{ \begin{array}{ll} 1 + q_{1}(z-1) + \frac{1}{2}\;
q_{2}\;(z-1)^{2},  & \mbox{model {\bf t},}  \\
\mbox{}  & \mbox{}  \\
\left[1 + q_{1}(1-z)\right]^{-1}, & \mbox{model {\bf g}.}
\end{array} \right.
\end{equation}
It is clear that the first case corresponds to the zero-one-two,
while the second to the geometric distribution of the offspring
number $\nu$.

\subsubsection{Survival probabilities}

By using Eq. (\ref{37}) in the model {\bfseries t} we obtain
\begin{equation} \label{46}
\frac{dS}{dx} = -(1-q_{1})\;S  - \frac{1}{2}\;q_{2}\;S^{2},
\end{equation}
and taking into account the initial condition $S(0)=1$ we have the
solution in the form
\begin{equation} \label{47}
S(x) = \left\{ \begin{array} {ll} e^{-(1-q_{1})x} \left[1 +
\frac{q_{2}}{2(1-q_{1})}\;(1 -
e^{-(1-q_{1})x})\right]^{-1}, & \mbox{if $q_{1}<1$,} \\
\mbox{} & \mbox{} \\
\frac{2}{2 + q_{2}\;x}, & \mbox{if $q_{1}=1$,} \\
\mbox{} & \mbox{} \\
2\;\frac{q_{1}-1}{q_{2}}\;\left[1 + (1 -
2\;\frac{q_{1}-1}{q_{2}})\;e^{-(q_{1}-1)x}\right]^{-1}, & \mbox{if
$q_{1}>1$}. \end{array} \right.
\end{equation}

The corresponding equation in the model {\bfseries g}  (i.e. when
the distribution of $\nu$ is geometric) is given by
\begin{equation} \label{48}
\frac{dS}{dx} = - \frac{1}{1 + q_{1} S} - S + 1 = - S\;\frac{1
-q_{1}(1-S)}{1 + q_{1}S},
\end{equation}
the solution of which can be easily obtained  in inverse form
\begin{equation} \label{49}
x(S) = \left\{ \begin{array}{ll} \frac{1}{1-q_{1}}\;\log
\frac{1}{S} + \frac{q_{1}}{1-q_{1}}\;\log[1 - q_{1}(1-S)], &
\mbox{if $q_{1}
\neq 1$,} \\
\mbox{} & \mbox{} \\
\frac{1-S}{S} + \log \frac{1}{S}, & \mbox{if $q_{1} = 1$,}
\end{array} \right.
\end{equation}
satisfying the initial condition $S(0)=1$.

The density function of the lifetime measured in $Q^{-1}$ units
can be calculated by using the relation
\[ \ell(x) = \frac{dL(x)}{dx} = - \frac{dS(x)}{dx}. \]
It is elementary to show that the density function is decreasing
monotonously from $\ell(0) = f_{0}$ to zero.  In Fig. $6$ one can
see the density function curves versus time $x=Qt$ for
subcritical, critical and supercritical trees and for both
distributions of $\nu$.
\begin{figure} [ht!]
\protect \centering{\includegraphics[height=12cm,
width=10cm]{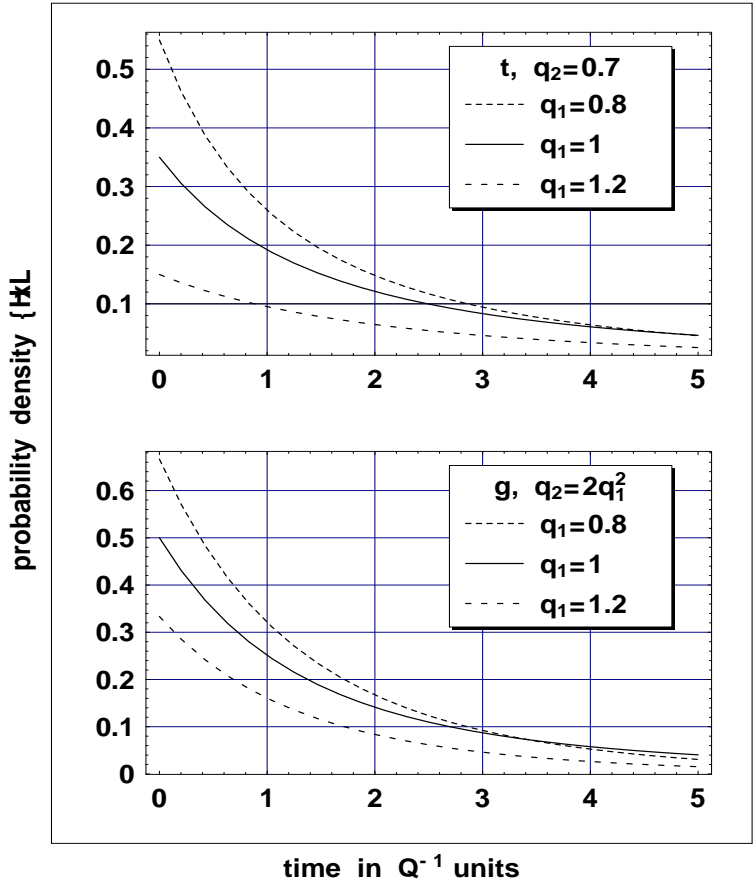}}\protect \vskip 0.2cm \protect
\caption{{\footnotesize Time dependence of the density function
$\ell(x)$ of the tree lifetime $\vartheta$ in models {\bfseries t}
and {\bfseries g}.}} \label{fig6}
\end{figure}

Let us calculate now the characteristic function of the random
variable $\vartheta = Q \theta$, i.e. of the tree lifetime
measured in $Q^{-1}$ units. Since the moments ${\bf
E}\{\vartheta^{j}\}, \;\; j=1,2,\ldots$ do not exist if the $q_{1}
\geq 1$, the calculations are restricted to the case when $q_{1} <
1$. One can write
\[ \varphi(\omega) = {\bf E}\{e^{-\omega \vartheta}\} =
\int_{0}^{\infty} e^{-\omega x}\;dL(x) =  - \int_{0}^{\infty}
e^{-\omega x}\;dS(x) = \int_{0}^{1} e^{-\omega x(y)}\;dy, \] where
$\omega$ is a complex number with $\Re \omega \geq 0$. Performing
the substitution
\[ x(y) = - \frac{1}{1-q_{1}}\;\log y\;\frac{1 +
2\frac{1-q_{1}}{q_{2}}}{y + 2\frac{1-q_{1}}{q_{2}}} \] in the
model {\bfseries t} one obtains the characteristic function
\begin{equation} \label{50}
\varphi_{t}(\omega) = (1 + \gamma)^{\omega \beta}\;\int_{0}^{1}
\left[\frac{y} {y + \gamma}\right]^{\omega \beta}\;dy,
\end{equation}
where \[ \beta = (1-q_{1})^{-1}\;\;\;\;\;\; \mbox{and}
\;\;\;\;\;\; \gamma = 2\frac{1-q_{1}}{q_{2}}, \;\;\;\;\;\; q_{1} <
1. \] In the model {\bfseries g} we should substitute for $x$ the
expression given by (\ref{49}), i.e. \[ x(y) =
\frac{1}{1-q_{1}}\;\log \frac{1}{y} +
\frac{q_{1}}{1-q_{1}}\;\log[1 - q_{1}(1-y)], \] and we find
\begin{equation} \label{51}
\varphi_{g}(\omega) = \int_{0}^{1}(1 - y)^{\omega \beta}\;(1 -
q_{1} y)^{\omega (1-\beta)} \;dy, \;\;\;\;\;\; q_{1} < 1.
\end{equation}

\subsubsection{Expectation and variance of the lifetime}

From the characteristic functions $\varphi_{t}(\omega)$ and
$\varphi_{g}(\omega)$ it can be easily calculated both the
expectation value and the variance of the lifetime $\vartheta$ of
a subcritical tree. In the model {\bfseries t} for the expectation
value one has
\[ {\bf E}\{\vartheta_{t}\} = -
\left(\frac{d\varphi_{t}(\omega)}{d\omega}\right)_{\omega=0} =
\beta\;\int_{0}^{1}\;\log \frac{y+\gamma}{y\;(1+\gamma)}\;dy =
\]
\begin{equation} \label{52}
\beta\;\gamma\;\log\left(1 + \frac{1}{\gamma}\right) =
\frac{2}{q_{2}}\;\log\left(1 +
\frac{1}{2}\;\frac{q_{2}}{1-q_{1}}\right).
\end{equation} while in the model {\bfseries g} one finds
expression
\begin{equation} \label{53}
{\bf E}\{\vartheta_{g}\} = -
\left(\frac{d\varphi_{g}(\omega)}{d\omega}\right)_{\omega=0} = 1 -
\log(1-q_{1}).
\end{equation}

\begin{figure} [ht!]
\protect \centering{\includegraphics[height=8cm,
width=12cm]{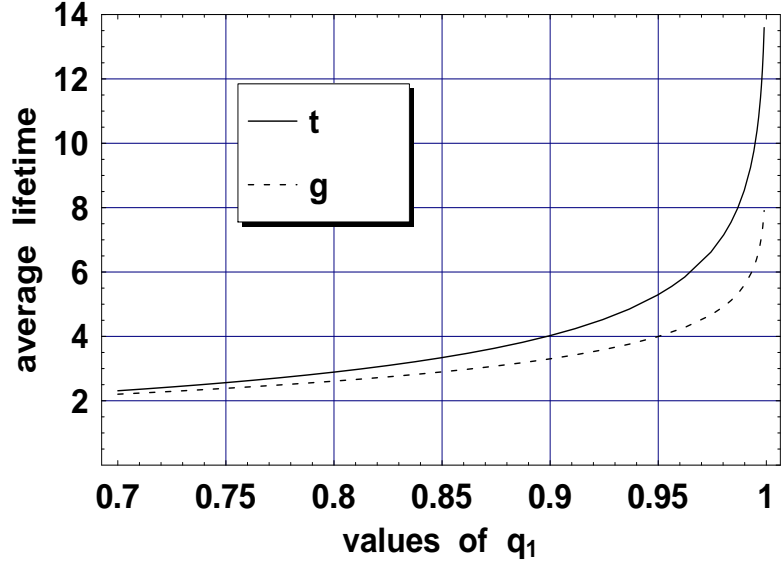}}\protect \vskip 0.2cm \protect \caption{
{\footnotesize Dependence of the average value of the tree
lifetime $\vartheta$ on $q_{1} < 1$ in models {\bfseries t} and
{\bfseries g}.}} \label{fig7}
\end{figure}

The curves in Fig. \ref{fig7} show the dependence of the
expectation of the tree lifetime on the average branching
parameter $q_{1}$ in both models {\bfseries t} and {\bfseries g}.
One can observe the difference between the two curves to be
unimportant. In both cases if $q_{1} \Rightarrow 1$ then the
expectation value becomes infinite like $\log (1-q_{1})^{-1}$.

For the calculation of the variance of the tree lifetime we need
the second moment of $\vartheta$ which can be immediately obtained
from the characteristic function. In the case of the model
{\bfseries t} we have
\[ {\bf E}\{\vartheta_{t}^{2}\} =
\left[\frac{d^{2}\varphi_{t}(\omega)}{d\omega^{2}}\right]_{\omega=0}
=  \beta^{2}\; \int_{0}^{1}\left[ \log
\frac{y+\gamma}{y\;(1+\gamma)}\right]^{2}\;dy. \] By using some
well known integral relations it can be shown that
\[ {\bf E}\{\vartheta_{t}^{2}\} = - \beta^{2}\;\gamma \left\{
\left[\log \left(1 + \frac{1}{\gamma}\right)\right]^{2} + 2 \;
Li_{2}\left(-\frac{1}{\gamma}\right)\right\}, \] where
$\;\;Li_{2}(u) = \sum_{k=1}^{\infty} u^{k}/k^{2}\;\;$
\begin{figure} [ht!]
\protect \centering{\includegraphics[height=12cm,
width=10cm]{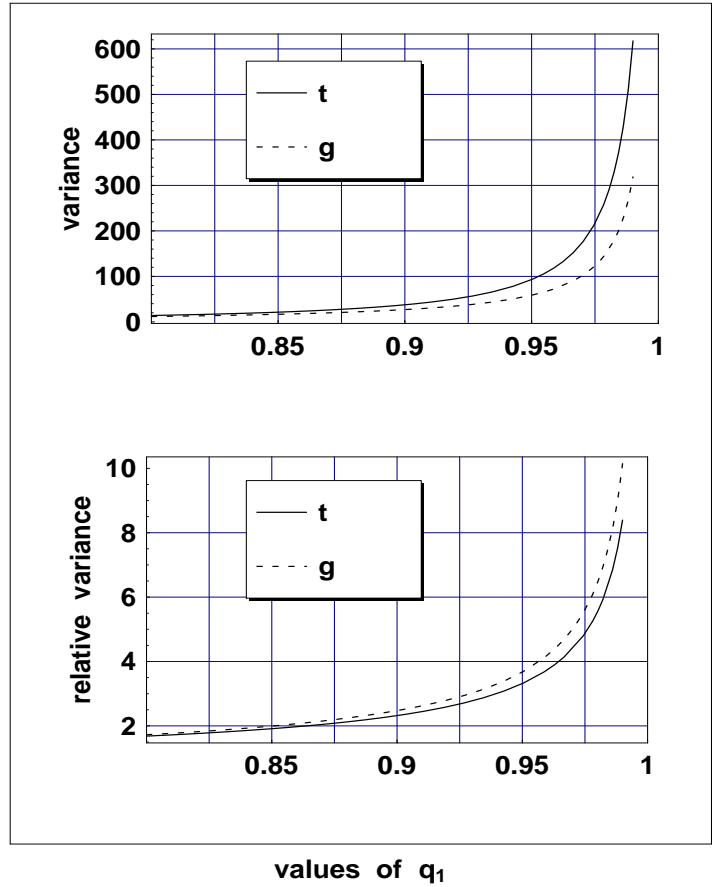}}\protect \vskip 0.2cm \protect
\caption{{\footnotesize Dependence of the variance and the
relative variance of the tree lifetime $\vartheta$ on $q_{1}<1$ in
models {\bfseries t} and {\bfseries g}.}}  \label{fig8}
\end{figure}
is the so called Jonqui\`ere's (dilogarithm) function. Taking into
account this form of the second moment we can write
\[{\bf D}^{2}\{\vartheta_{t}\} = -
\left(\frac{2}{q_{2}}\right)^{2}\;\left(1 +
\frac{1}{2}\;\frac{q_{2}}{1-q_{1}}\right)\;\left[\log\left(1 +
\frac{1}{2}\;\frac{q_{2}}{1-q_{1}}\right)\right]^{2} - \]
\begin{equation} \label{54}
\frac{4}{q_{2}\;(1-q_{1})}\;Li_{2}\left(-\frac{1}{2}\frac{q_{2}}{1-q_{1}}\right).
\end{equation}
Performing similar calculations in the case of the model
{\bfseries g} for the variance of $\vartheta_{g}$ we obtain
\begin{equation} \label{55}
{\bf D}^{2}\{\vartheta_{g}\} = 1 - \frac{1}{1-q_{1}}\;\left[\log
(1-q_{1})\right]^{2} -
\frac{2}{1-q_{1}}\;Li_{2}\left(-\frac{q_{1}}{1-q_{1}}\right).
\end{equation}

In Fig. \ref{fig8} one can see the dependence of the variance as
well as the relative variance of the tree lifetime on the
parameter $q_{1}$ in the cases of both models {\bfseries t} and
{\bfseries g}. When $q_{1}$ is approaching to $1$ from below the
fluctuation of the tree lifetime becomes unlimitedly large, and so
in the vicinity of the critical state the average lifetime loses
almost completely its information content.

\section{Concluding remarks}

The probability distribution of the number of nodes with $k \geq
0$ outgoing lines has been investigated in randomly evolving trees
defined in \cite{lpal02}. Special attention was paid on the
stochastic properties of end-nodes. We found that the birth and
death of end-nodes in randomly evolving trees are playing a
decisive role in the dynamics of the process.

It is remarkable that the {\em relative variance} of the number of
end-nodes vs. time has well-defined maximum when the evolution is
either subcritical or supercritical. In the case of critical
evolution the relative variance increases monotonously with the
time. On the contrary, the {\em relative dispersion} of the number
of dead end-nodes vs. time has a minimum just after the beginning
of the evolution. The minimum can be seen in each of evolution
states.

We defined the lifetime of randomly evolving trees and derived a
non-linear differential equation for the probability that the
lifetime is larger than a given positive real number $x$. Three
theorems have been proven to obtain asymptotic expressions for the
survival probability of subcritical, critical and supercritical
trees. Though the average lifetime of supercritical trees is
always infinite it has been shown that the probability of finite
lifetime of supercritical trees is larger than zero. In other
words, randomly evolving supercritical trees may have finite
lifetime.

\end{document}